\documentstyle[aps,prl,psfig]{revtex}

\begin{document}
\draft

\title{Conductance suppression in normal metal-superconductor
 mesoscopic strucutres }

\author{C.J. Lambert$^*$,R. Seviour$^*$ and A.F.Volkov$^{\dagger*}$}
\address{$^*$ School of Physics and Chemistry,
Lancaster University, Lancaster LA1 4YB, U.K.\\
$^{\dagger}$Institute of Radioengineering and Electronics of the Russian
   Academy of Sciencies, Mokhovaya str.11, Moscow
   103907, Russia}
\date{\today}
\maketitle
\begin{abstract}
 Using a scattering matrix approach and quasiclassical Green's function 
technique, we calculate the conductance of the S/N system (see Fig.1).
We establish that the difference between the superconducting and normal 
state conductance $(\delta G = G_s - G_n)$ is negative for large S/N 
interface resistances $(R_{S/N})$ and changes sign with decreasing $R_{S/N}$. 
The comparision of the results obtained with experimental data is carried out.
\end{abstract}


{\bf I. Introduction}
\smallskip

 Recent studies of transport properties of mesoscopic, normal metal-superconductor 
(N/S) structures (see
reviews \cite{r1,r2}),have revealed a number of new physical phenomena. 
Examples include 
the subgap conductance measured of SIN (superconductor-insulator-normal metal) 
junctions \cite{r3,r4,r5,r6}, oscillations in the magneto-
conductance of N/S systems with normal or superconducting loops 
\cite{r7,r8,r9,r10,r11,r12},and the nonmonotonic dependence 
of the conductance on temperature and voltage \cite{r10,r13}. Although the majority
of experimental results have been successfully explained, there are
some which remain anomalous. In particular the increase 
in resistance of diffusive N/S systems in a certain temperature range below $T_{c}$ 
\cite{r14,r15,r16} has remained unexplained for a number of years. An early theoretical 
prediction \cite{a} that superconductivity induced conductance suppression is a 
generic feature of N/S nanostructures was followed by quantitative theories 
of this effect in the ballistic and Anderson localisied regions \cite{b} as 
well as in resonant structures \cite{c}. However a quantitative theory in the 
diffusive region has remained elusive. Several authors have suggested  
possible explanations of this puzzling phenomenon. In the simplest, 
 \cite{r14}, the resistance change $\delta R_{s}= R_{s}- 
R_{n}$ is determined by a change in the interface resistance $R_{b}$ which 
is larger in the superconducting state than in the normal state. 
Another possibility is presented in \cite{r17,r18} where a two dimensional, 
multiprobe geometry was considered. The current $I_{a}$ passes through 
two contacts on one side of the normal film contacting a superconductor 
and the voltage $V_{b}$ is measured between two probes located on the 
opposite side of the normal film. The authors of Ref\cite{r17,r18} showed that the 
quantity $R_{ab} = V_{b}/I_{a}$ may  exhibit an increase below $T_{cs}$ compared with its 
normal state value. In this geometry the spatial distribution of the current is 
nonuniform. However in some experiments, the geometry is almost one 
dimensional with the current distribution across the width of the normal film 
almost uniform. Therefore this mechanism may not be responsible for 
all the experimental observations of enhanced resistance.

 In this paper we suggest an alternative mechanism which determines the change 
in resistance $\delta R_{s}$ (or the conductance  $\delta G_{s} \approx -  
\delta R_{s}/ R^{2}_{n}$) of the structure shown in Fig. 1. We will show that  
$\delta R_{s}$ may be positive if the interface resistance $R_{b}$ is large 
enough compared with the resistance of the metallic film in the normal state 
$R_{n}$. The variation  $\delta R_{s}$  is 
determined by two factors: a variation of the shunting interface 
resistance $\delta R_{b}$, leading to a positive change in resistance 
and a variation of the normal film resistance due to a condensate induced 
by the proximity effect. In view of these conflicting effects it is not obvious 
what sign $\delta R_{s}$ will adopt for any given parameters of the system. 
In what follows we use two methods to study the change in resistance, namely an 
analytical quasiclassical technique and a numerical scattering approach. 
The scattering approach \cite{r2} complements the quasiclassical method, and 
enables us to probe areasof parameter space which lie outside the region of 
validity of the latter. 

\smallskip
{\bf II. Quasiclassical Theory}
\smallskip

 Consider a diffusive regime where the mean free path is shorter than any 
other characteristic length in the system (except the Fermi wavelength). 
Such a case is realized in most experiments performed on metallic films or 
on doped semiconductors. For diffusive S/N mesoscopic structures,  
equations for the quasiclassical Green's functions were derived 
many years ago and are presented (in the most convenient form suitable for the 
present analysis ) in Larkin and Ovchinnikov's paper \cite{r19}. These 
equations must be supplemented by  boundary conditions at the S/N interface 
derived by Zaitsev \cite{r20} (see also \cite{r21,r22}) and have been used 
extensivly for the theoretical study of transport properties of S/N mesoscopic 
structures \cite{r23,r24,r25,r26,r27,r28,r29,r30,r31,r32,r33}.

 In this paper we shall assume that the proximity effect is weak, i.e. the 
amplitude of the condensate induced in the normal film is small. We will 
show that this is true for structures where the S/N interface resistance ($R_{b}$) 
, in the normal state, exceeds the resistance of the normal film. When this 
condition is satisfied, the condensate functions obey the linearized Usadel 
equation and the distribution function obeys the equation (see \cite{r29,r30}), 

\begin{equation}
 L^{2} \partial_{x} ((1-m)\partial_{x}f) = f g_{b} G_{b} \vartheta (x \in (S/N)),
\label{eq1}
\end{equation}

 where the function $\vartheta (x)$ is equal to 1 in the S/N region and zero 
otherwise, $m=\frac{1}{8} Tr( \hat{F}^{R} - \hat{F}^{A})^{2}$ and $\hat{F}^{R(A)}$
 is the retarded (advanced) Green's function, $g_{b} = \rho L^{2} /R_{b \Box} 
d = g_{b1} (L/L_{1})$ , $g_{b1}$ is the ratio of the normal film resistance to the 
S/N resistance, $R_{b \Box}$  is the S/N imterface resistance per unit area in 
the normal state, $\rho$ and $d$ are the specific resistivity and the thickness of 
the normal film. The function $G_{b}(x)$ determines the local normalized 
conductance of the S/N interface in the superconducting state,

\begin{equation}
G_{b}(x) = \nu_{s} \nu_{n} +\frac{1}{8}Tr(\hat{F}^{R} + \hat{F}^{A})(\hat{F}^{R}_{s}
 + \hat{F}^{A}_{s}),
\label{eq2}
\end{equation}

 where the density of states (DOS) in the superconductor
$\nu_{s} = Re (\epsilon + i \Gamma)/\surd((\epsilon + i \Gamma)^{2} -\Delta^{2})$
 and $\nu_{n}$ is the DOS in the normal film (for simplicity we assume $\nu_{n}= 1$ 
,i.e. the S and N metals are regarded as identical apart from the critical 
temperature, we also assume that $T_{cn}=0$).The first term in Eq. (\ref{eq2}) 
describes the contribution of the quasiparticle current to the conductance 
(if $\Gamma=0$ it differs from zero only at energies $|\epsilon|>\Delta$). 
The second term is due to Andreev reflection and discribes a conversion of 
the low-energy quasiparticle current into the condensate current
(if $\Gamma=0$, the current is not zero for $|\epsilon|<\Delta$). The condensate 
functions $\hat{F}^{R(A)}_{s}$ in the superconductor are assumed undisturbed by the 
proximity effect (this is true provided that $\Delta \gg \gamma_{b} = K^{2}_{b} D$
, where $K^{2}_{b} = (R_{b \Box} d \sigma)^{-1}$), and they are equal to,

\begin{equation}
\hat{F}^{R(A)}_{s} = i\hat{\tau}_{y} F^{R(A)}_{s},
\label{eq3}
\end{equation}

 where $F^{R(A)}_{s}=\Delta/\surd((\epsilon + i \Gamma)^{2} -\Delta^{2})$,
$\Gamma$ is the damping rate in the excitation spectrum of the 
superconductor. Assuming that the right-hand side of Eq. (\ref{eq1}) is a 
small perturbation we easily find a solution for,

\begin{equation}
f=\left\{
\begin{array}{cc}
\left( J_{1} [ x + \int^{x}_{0} dx_{1} (m + m_{b})]  \right) , & 0 < x < L_{1} \\
\left( J(x-L_{1}) + f(L_{1})  \right), & L_{1} < x < L
\end{array}
\right.
\label{eq4}
\end{equation}

 where $J_{1}$ and $J$ are the energy dependent intergration constants. The current 
$I$ through the system is expressed in terms of $J$ ( see \cite{r29,r30}),

\begin{equation}
I = (\sigma d/2e) \int d\epsilon J(\epsilon).
\label{eq5}
\end{equation}
 The function $m_{b}$ in Eq.(\ref{eq4}) is given by,

\begin{equation}
m_{b}= (g_{b}/L^{2}) \int^{x}_{0} dx_{1} x_{1} G_{b}(x_{1}).
\label{eq6}
\end{equation}

 In the reservoirs the distribution function has the equilibrium form, 

\begin{equation}
f(\epsilon,L)=F_{v}=\left[\tanh \left(\epsilon +eV\right)\beta 
-\tanh  \left(\epsilon
-eV\right)\beta \right]/2,
\label{eq7}
\end{equation}

 where $\beta=(2Tk_{B})^{-1}$. By matching the functions and their derivatives 
at $x=L_{1}$ and using Eq. (\ref{eq7}), we find for the "partial current" 
$J(\epsilon)$,

\begin{equation}
J(\epsilon)=(F_{v}/L)[1-<m>+l_{1}(m_{b1}-<m_{b}>_{1})],
\label{eq8}
\end{equation}

 The angle brackets mean a spatial averaging over the regions (0,L) 
and $(0,L_{1}); m_{b1} = m_{b} (L_{1}), l_{1} = L_{1}/L$. With the aid of Eqs. 
(\ref{eq5}) and (\ref{eq8}), we find the normalized difference between the differential 
conductances of the system in the superconducting state and the normal state,

\begin{equation}
\delta S = \frac{ G_{s} -  G_{n}}{ G_{n}} = - \int^{\infty}_{0}  d\epsilon 
\beta F_{v}^{'}
[<m>+l_{1}(<m_{b}-m_{b}^{n}>_{1} - (m_{b1}-m_{b1}^{n}))]
 / [ 1 + g_{b} l^{3}_{1}/3],
\label{eq9}
\end{equation}

 Here $G_{s,n} = (dI/dV)_{s,n}$ is the differential conductance below and 
above $T_{cs}$; $m_{b}^{n}$ is the function $m_{b}(x)$ in the normal state:
$m_{b}^{n} = g_{b} x^{2} / 2 L^{2}$, $F_v'=\partial F_v/\partial(eV{\beta})$.

 Let us discuss the physical meaning of the different terms in Eq. (\ref{eq9}).
The first term gives a positive contribution to $\delta S$ ($<m>$ is negative). 
This term arises from the renormalization of the normal film conductance caused 
by the induced condensate. This has been calculated in several papers 
\cite{r30,r31,r32,r33}, where it was established that this term 
has a nonmonotonic voltage and temperature dependence decreasing to 
zero at $[eV,T]=0$ and $[eV,T] \gg \epsilon_{L}$ ($\epsilon_{L} = D / L^{2}$ 
is the Thouless energy). The second term in Eq. (\ref{eq9}) 
determines the change in conductance due to different values of the 
S/N interface resistance in the normal and superconducting 
states. The contribution of this term to  $\delta S$ is negative 
because the S/N interface resistance in the superconducting state 
is larger than the normal state interface resistance (as long as 
the barrier transparancy is not too high). Let us estimate the magnitudes 
of these terms. If $g_{b}$ is small, the second term in 
Eq. (\ref{eq2}) (the subgap conductance) is small compared 
with the first term. Therefore at low temperatures the 
contribution caused by the second term in Eq. (\ref{eq9}) 
is related to a change in the DOS of the superconductor. This yields,

\begin{equation}
\delta S_{dos} \approx - g_{b} l_{1}^{3} / 3.
\label{eq10}
\end{equation}

 As we shall see, the amplitude of the condensate functions $\hat{F}^{R(A)}$ 
induced in the normal film by the proximity effect is of the order $g_{b} l_{1}$ 
, i.e. of the order of the ratio of the normal film and the S/N 
interface resistances. The characteristic energy of decay of  
$\hat{F}^{R(A)} (\epsilon)$ is the Thouless energy $\epsilon_{L}$.
Thus the contribution to $\delta S$ from the proximity effect 
(the first term in Eq. (\ref{eq9})) is (if $\epsilon_{L} <T \ll 
\Delta$ and $l_{1} \ll 1$),

\begin{equation}
\delta S_{pr} \approx  g_{b}^{2} l_{1}^{2} .
\label{eq11}
\end{equation}

 Comparing Eqs. (\ref{eq10}) and (\ref{eq11}), we see that the sign of 
 $\delta S$ changes from negative to positive as $g_{b}$ increases and 
$\delta S_{min} \approx -l_{1}^{4}$ is reached when $g_{b} \approx l_{1}$. 

 In order to find  $\delta S$, we need to calculate $\hat{F}^{R(A)}$. 
As noted above, in the limit of a weak proximity effect the function 
$\hat{F}^{R(A)}$ obeys the linearized Usadel equation which 
may be presented in the form (see for example \cite{r29,r30}),

\begin{equation}
 \partial _{xx} \hat{F}^{R(A)} - \left(k^{R(A)}\right)^2\hat{F}^{R(A)}=
 - (g_{b} /L^{2}) \hat{F}_{s}^{R(A)} \vartheta (x \in (S/N)),
 \label{eq12}
\end{equation}

where $(k^{R(A)}_{b})^{2}=(\mp 2i\epsilon + \gamma)/D$, $\gamma$ is the 
depairing rate in the normal film. The solution to Eq. (\ref{eq12}) 
satisfying the boundary conditions $\hat{F}^{R(A)}(\pm L) = 0$ is 
the function,

\begin{equation}
\hat{F}=(g_{b} /\theta^{2})\hat{F}_{s} \left\{
\begin{array}{cc}
\left( 1-c_{2} \cosh(kx)  \right) , & 0 < x < L_{1} \\
\left( s_{1} \sinh(k(L-x))  \right), & L_{1} < x < L
\end{array}
\right.
\label{eq13}
\end{equation}

 Here $\theta = kL$, $c_{2}=\cosh(\theta_{2})/\cosh(\theta)$,$ s_{1}=\sinh(\theta_{1})/\cosh(\theta)$,
$\theta_{1,2} = k L_{1,2}$, $L_2=L-L_1$. For convenience we have dropped the 
indices R(A). One can see from Eq. (\ref{eq13}) that at characteristic 
energies $\epsilon \approx\epsilon_{L} \ll \Delta$ the amplitude 
of $\hat{F}^{R(A)}$ is of the order $g_{b} l_{1}$. It is worth 
noting that if $ l_{1} \ll 1$ (this condition is satisfied in most 
experiments), then the magnitude $g_{b} l_{1}$ corresponding to the actual 
$g_{b}$ is of the order $l_{1}^{2}$,i.e. as $l_{1}$ is small, the 
proximity effect is small.

 From Eq.(\ref{eq13}), we can find the quantites $<m>$,$m_{b1}$ and $<m_{b1}>$. 
Substituting them into Eq. (\ref{eq9}), we obtain the variation 
of the normalized conductance $\delta S$ as a function of temperature,$g_{1}$,
$l_{1}$ etc (see Appendix). In the general case the expression for 
$\delta S$ has a rather complicated form, but may be simplified 
drastically by taking the zero-bias, zero-temperature limit, 
in which case $\delta S$ becomes,
\begin{equation}
\delta S_{0} = g_{b} l_{1}^{3} [g_{b}l_{1}(1 - l_{1}\frac{4}{15}) -\frac{1}{3}].
\label{eq14}
\end{equation}

 In this case the contribution due to the variation of the normal 
region conductance goes to zero (the term $<m>$ in Eq.(\ref{eq9})). 
The variation $\delta S$ is caused by a change in the S/N interface 
conductance $G_b$; the first term in Eq.(\ref{eq2}) gives a negative 
contribution, and the second term (i.e. the subgap conductance) gives a 
positive contribution. This expression also changes sign at 
$g_{b}l_{1}\approx 1/3$. However in this case the condensate amplitude 
is not small,and strictly speaking, Eq. (\ref{eq14}) is not valid 
when $g_{b}l_{1}$ is of order 1. Nevertheless, we show 
later using numerical calculations that this conclusion regarding the 
change in sign of $\delta S$ remains valid in the 
zero-bias, zero temperature limit. Fig. 2 shows 
the dependence $\delta S(g_{b})$ for $l_{1} = 0.2$ and 
$l_{1} = 0.4$. We see that in accordance with qualitative 
speculations given above, $\delta S$ is negative at small 
$g_{b}$, reaching a minimum with increasing $g_{b}$ and 
then changing sign. The magnitude of $\delta S_{min}$ 
decreases with decreasing $l_{1}$ and depends on 
temperature in a complicated nonmonotonic way (see Fig. 3).

 In Fig. 4 we show the dependence of $\delta S$ on $\gamma$ 
($\gamma$ may increase by applying an external magnetic field, 
$\gamma\sim H^2$), the effect of a negative $\delta S$ 
becomes more pronounced. This behaviour is quite clear from a 
physical point of view. The applied magnetic field suppresses 
the proximity effect, but affects the DOS only weakly. Therefore 
the relative contribution $\delta S_{dos}$ increases with increasing magnetic 
field. A similar situation takes place in fluctuation 
paraconductivity in layered superconductors where an increase 
in the resistance due to superconducting fluctuations is 
enchanced by the magnetic field \cite{r34}.

 In Fig. 5 we plot the voltage dependence of 
$\delta S$. One can see that two maxima exist in this 
dependence; one is close to zero bias and another 
one located at $eV\approx\Delta$. A similar voltage 
dependence of the phase coherent conductance has been observed 
in the recent work of \cite{r35}. We note that although the effect 
of the conductance decrease is small ($\leq 5\%$), in dimensional units 
the conductance decrease $\delta G$ may be much larger than 
the quantum conductance $2e^2/h$ ( $\delta G \gg 2e^2/h$). 
Only in this limit can we use quasiclassical theory.

\smallskip
{\bf III. Numerical simulations}
\smallskip

 In this section we use the scattering approach reviewed in \cite{r2} to 
determine $\delta S$ ,for a tight binding lattice with the geometry of Fig. 1. 
In the linear-response limit, at zero temperature, the conductance of a 
phase-coherent structure may be calculated from the fundamental current 
voltage relationship \cite{c18},

\begin{equation}\left(\matrix{I_1\cr  I_2}\right)=
\left(\matrix{a_{11}&a_{12}\cr a_{21}&a_{22}}\right)
\left(\matrix{v_1- v\cr  v_2- v}\right)
\label{n4},
\end{equation}
where, at finite temperatures,

\begin{equation}\left(\matrix{a_{11}&a_{12}\cr a_{21}&a_{22}}\right)=\frac{2e^2}{h}
\int^\infty_{0}\, dE\,(-{{\partial f(E)}\over{\partial E}})\left(\matrix{N^+_1(E)- R_0(E)+R_a(E)
&T^\prime_a(E)-T^\prime_0(E)\cr T_a(E)-T_0(E)&N^+_2(E)-R^\prime_0(E)+R^\prime_a(E)}\right)
\label{n5},\end{equation}

Eq. (\ref{n4}) relates the current $ I_i$ 
from a normal reservoir $i$ to the voltage differences $(v_j - v)$, 
where $v=\mu/e$ ($\mu$ is the chemical potential of the superconductor).
The $a_{ij}$'s are linear combinations of normal $(T_0,R_0)$ and Andreev $(T_a,R_a)$ 
scattering coefficients. The primes on the coefficients refer to quasiparticles originating 
from the right-hand reservoir, whilst the coefficients without primes refer to particles 
from the left reservoir. Setting $I_1=I=-I_2$ and solving Eq. (\ref{n4}) the 
two probe conductance  is (see \cite{c20}),

\begin{equation}
G=\frac{a_{11}a_{22}-a_{12}a_{21}}{ a_{11}+a_{22}+a_{12}+a_{21}}
\label{a24}.
\end{equation}

As noted in \cite{c20} in the presence of disorder, the various transmission 
and reflection coeffcients can be computed by solving the
Bogoliubov - de Gennes equation on a tight-binding lattice of
sites, each labelled by an index $i$ and possessing
a particle (hole) degree of freedom $\psi(i)$ $(\varphi(i))$ 
($\psi (\varphi)$ is the particle (hole) wavefunction). 
In the presence of local s-wave pairing described by a
superconducting order parameter $\Delta_i$, this takes the form,

\begin{eqnarray}
\begin{array}{c c}
E\psi_i
=&\epsilon_i \psi_{i}
-\sum_{\delta} \tau \left(  \psi_{i+\delta} + \psi_{i-\delta} \right)
+ \Delta_{i} \varphi_{i}\\
E\varphi_i =&-
\epsilon_i \varphi_{i}
+\sum_{\delta} \tau  \left(  \varphi_{i+\delta}+ \varphi_{i-\delta}\right)
+\Delta^*_{i}\psi_{i}.\\
\end{array}
\label{2}
\end{eqnarray}

 In what follows, in the normal diffusive region, the on-site energy 
$\epsilon_i$ is chosen to 
be a random number,uniformly distributed over the interval $\epsilon_0 -1$ 
to $\epsilon_0+1$, whereas in the clean N-regions $\epsilon_i=\epsilon_0$. In 
the S-region, the order parameter is set to a constant, $\Delta_i=\Delta_0$, 
while in all other regions, $\Delta_i=0$. The nearest neighbour hopping element 
$\tau$ merely fixes the energy scale (i.e. the band-width), whereas $\epsilon_0$ 
determines the band-filling. In what follows we choose $\tau=1$.
By numerically solving for the scattering matrix of Eq. (\ref{2}),
exact results for the dc conductance can be obtained \cite{r18,c18,c19}. 
In the zero bias, zero temperature limit, Eq. (\ref{a24}) is greatly simplified 
and reduces to,

\begin{equation} 
G= T_0 + T_a + \frac{2( R_{a} R_{a}^{'}- T_{a} T_{a}^{'} )}
{ R_a + R_{a}^{'} + T_a + T_{a}^{'} }.
\label{a25}
\end{equation}

 For the structure shown in figure 1 , with a superconductor of length $2L_{1}$
, and a barrier resistance R, evaluation of this expression yields results for 
$<G_n>,<G_s>$ and $<\delta G>$ shown in table 1. In each case, the normal 
diffusive region is 40 sites wide and 64 sites long. The superconductor is of 
width 20 sites with $\Delta_0=0.1$ ($\Delta_0=0$) in superconducting (normal) 
state. Results are obtained by averaging over 100 disorder realizations, yielding 
an estimated error in the mean values of approximately 0.04.
the first row of the table shows results for $L_{1}=30,R=2$ and 
demonstrates that a negative $\delta G$ can indeed occur,
 and comparison with the $R=0.5$, shows that lowering the interface 
resistance causes $\delta G$ to change sign. As discussed previously, this 
result was expected although could not be proved using quasiclassical theory.
Also, as discussed previously, when $L_{1}$ is decreased (e.g. to $L_{1}=20$ 
with $R=2$) the table shows that $\delta G$ changes sign and becomes 
positive. Finally, to examine the effect of a magnetic field, the forth row 
of the table shows results with a magnetic field 
applied to the normal region (corresponding to 0.8 flux quanta through
the whole structure). This demonstrates that
 the introduction of a magnetic field causes a negative $\delta G$ to 
become more negative in agreement with the 
quasiclassical approach  but in conflict with experimental evidence.

\smallskip
\begin{tabular}{ccccc}
$L_{1}$&R&$<G_n>$&$<G_s>$&$<\delta G>$ \\[0.5ex]
30&2&3.70&3.40&-0.30\\
30&0.5&4.10&4.26&0.16\\
20&2&2.93&3.06&0.14\\
30&2&3.79&3.47&-0.33\\
\end{tabular}
{\bf Table 1}
\smallskip

 Finally we note that at a finite temperature ($k_{B}T=\epsilon_{L}$)
, where the full integral of Eq.(\ref{n5}) needs to be 
evaluated, we find for the structure of row 1 in the table,
$<G_n>=3.68$,$<G_s>=3.56$,$<\delta G>=-0.12$, which confirms the 
prediction made using the quasiclassical 
approach, that the onset of superconductivity causes a drop in the conductance 
of the structure, even at finite tempretures.

\smallskip
{\bf IV. Discussion}
\smallskip

 We have demonstrated that superconductivity-induced conductance 
suppression is an inherent property of the structure of figure 1. 
The suppression of the conductance at temperatures below $T_{cs}$, 
is $\leq 5\%$ and it is enhanced by the application of a magnetic 
field. In the experiment of \cite{r16} a stronger effect ($10-20\%$) is observed, 
which decreases when a rather weak magnetic field is applied. This 
suggests that the magnitude of the effect and the field dependence 
are geometry-dependent. For the future, it would be of interest to 
confirm this experimentally by measuring the conductance of 
S/N structures of the type shown in Fig. 1 
with different ratios of the normal channel and S/N 
interface resistances.

\smallskip
{\bf  V. Acknowledgements}
\smallskip

 One of us (A.F.V) is grateful to the Royal Society, to the 
Russian Fund for Fundamental Research (Project 96-02-16663 a) 
and to CRDF (project RP1-165) for financial support. We also 
wish to thank the EPSRC, for their financial support. 

\smallskip
{\bf VI. Appendix}
\smallskip

 Using Eq. (\ref{eq12}) we easily find the expression 
for the quantites in Eq. (\ref{eq8}). We have,

\begin{equation}
<m> = - \frac{1}{2}[|F_{s}|^2 <|b|^2> + Re(<b^2>F_{s}^2)],
\label{a1}
\end{equation}
where $F_{s}$ is defined in Eq.\ref{eq3},

\begin{equation}
<|b|^2> = \frac{g_{b}^2}{|\theta|^{4}} (l_{1} [1+\frac{|c_2|^2}{2}(
\frac{\sinh(2l\theta')}{2l_1\theta'}+\frac{\sinh(2l_1\theta'')}{2l_1\theta''})
- 2Re(c_2\frac{\sinh(l_1\theta)}{l_1\theta})]+l_2\frac{|s_1|^2}{2}(
\frac{\sinh(2l_2\theta')}{2l_2\theta'}-\frac{sin(2l_2\theta'')}{2l_2\theta''})),
\label{a2}
\end{equation}

\begin{equation}
<b^2> =  \frac{g_{b}^2}{\theta^{4}} (l_{1} [1+\frac{c_2^2}{2}(
\frac{\sinh(2l_1\theta)}{2l_1\theta}+1)
- 2c_2\frac{\sinh(l_1\theta)}{l_1\theta}]+l_2\frac{s_1^2}{2}(
\frac{\sinh(2l_2\theta)}{2l_2\theta}-1)),
\label{a3}
\end{equation}

\begin{equation}
m_{b1}-m_{b1}^n =  g_{b}l_{1}^2 [ \frac{\nu_s-1}{2}+\frac{g_{b}}{2} Re 
 \frac{A}{\theta^2}(\frac{1}{2}-c_2(\frac{\sinh(\theta l_1)}{\theta l_1}-
 \frac{\cosh(\theta l_1)-1}{(\theta l_1)^2})],
\label{a4}
\end{equation}

\begin{equation}
<m_{b}>_1-<m_{b}^n>_1 =  g_{b}l_{1}^2 [ \frac{\nu_s-1}{6}+\frac{g_{b}}{2} Re 
 \frac{A}{\theta^2}(\frac{1}{6}-\frac{c_2}{(\theta l_1)^2}(\cosh(\theta l_1)+1-
\frac{2\sinh(\theta l_1)}{\theta l_1}))],
\label{a5}
\end{equation}

$c_2=\frac{\cosh(\theta l_2)}{\cosh(\theta)}$,$s_1=\frac{\sinh(\theta l_1)}{\cosh(\theta)}$
,$\theta=kL$,$A=|F_s|^2 - F_s^2$, $\theta'= Re[\theta],\theta''= Im[\theta]$.

\newpage

{\bf VII. References}

\begin{figure}
\centerline{\psfig{figure=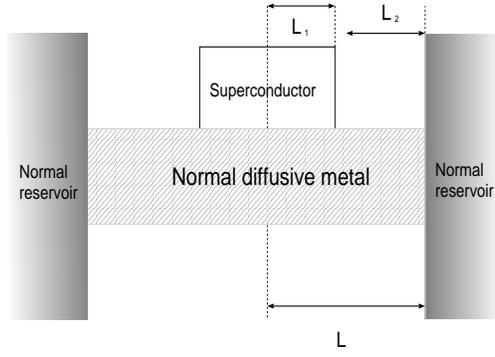,width=7 cm,height=6cm}}
\caption{The structure considered.}
\label{fig1}
\end{figure}

\begin{figure}
\centerline{\psfig{figure=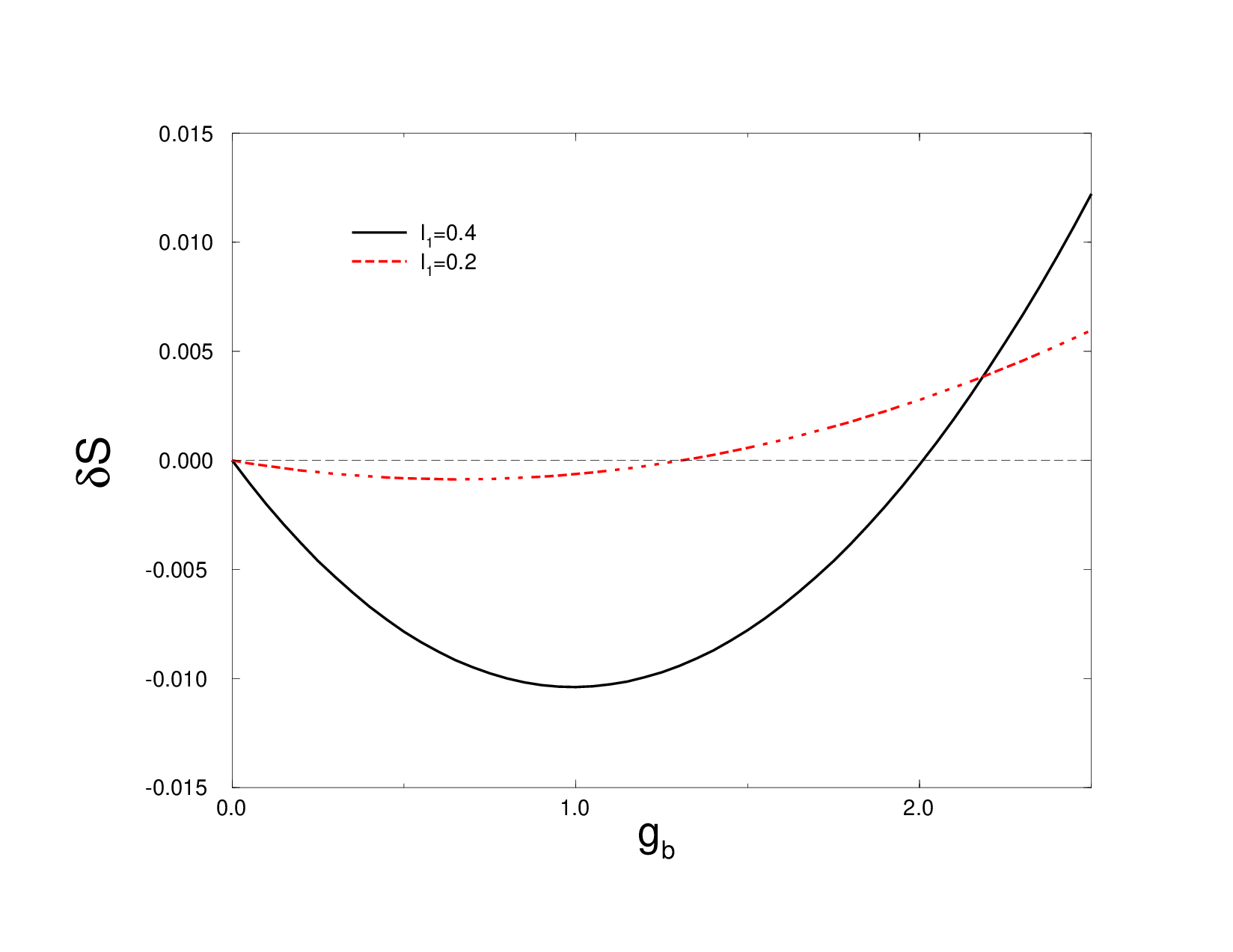,width=8 cm,height=6cm}}
\caption{The two curves show the dependence of $\delta S$ on  $g_{b}$, 
for the solid curve $l_1 = 0.4$ and for the dashed curve $l_1 =0.2$. Where both 
curves have the parameters $\Gamma = 0.1, \Delta = 10, 
\gamma = 0$ , $\alpha = 1.0$ and $V=0$. }
\label{fig2}
\end{figure}

\begin{figure}
\centerline{\psfig{figure=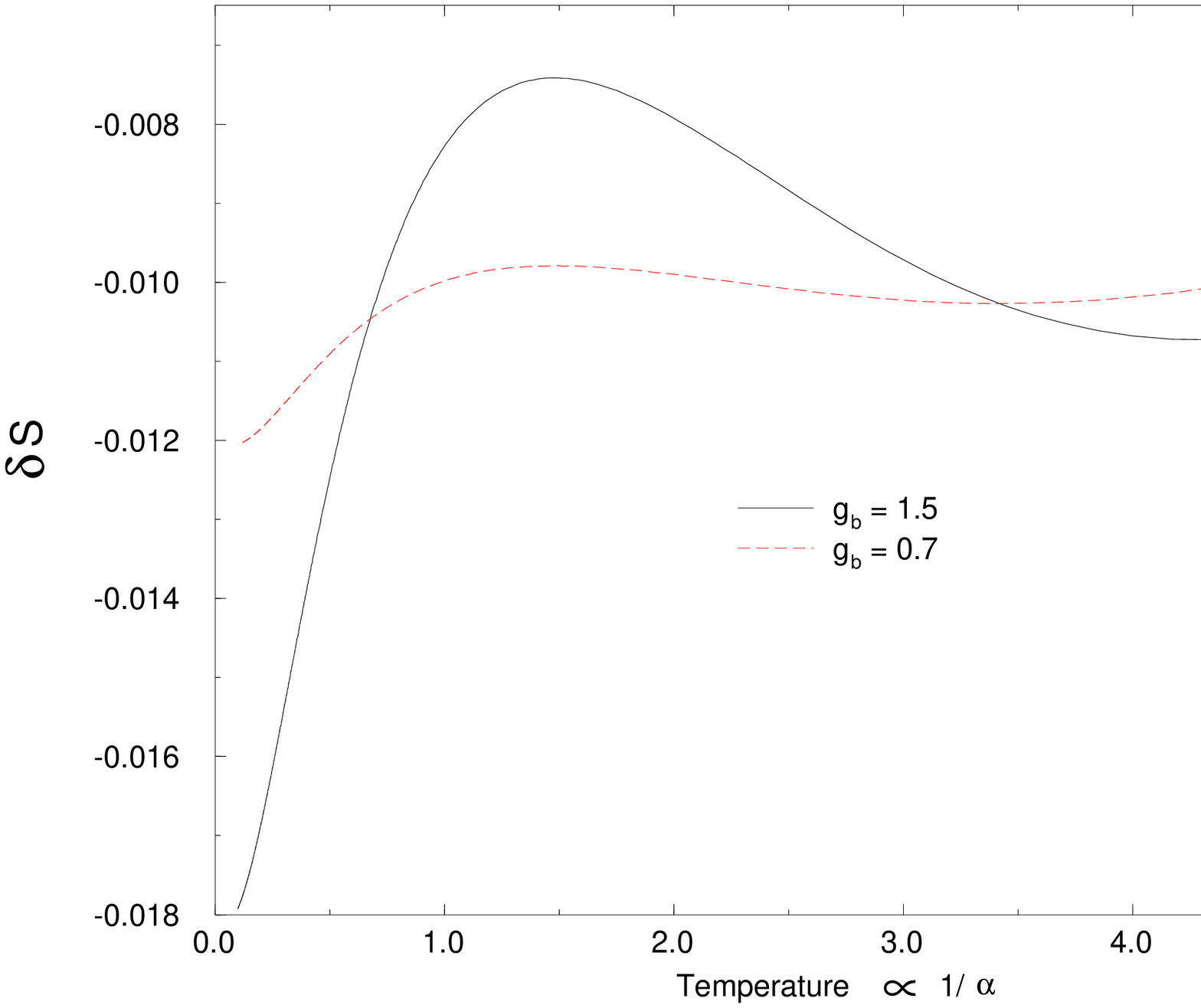,width=8 cm,height=6cm}}
\caption{Results showing dependence of $\delta S$ on temperature for $g_{b} 
= 1.5,0.7, \Gamma = 0.1, l_1 = 0.4, \Delta = 10, \gamma = 0$ and $V=0$. }
\label{fig3}
\end{figure}

\begin{figure}
\centerline{\psfig{figure=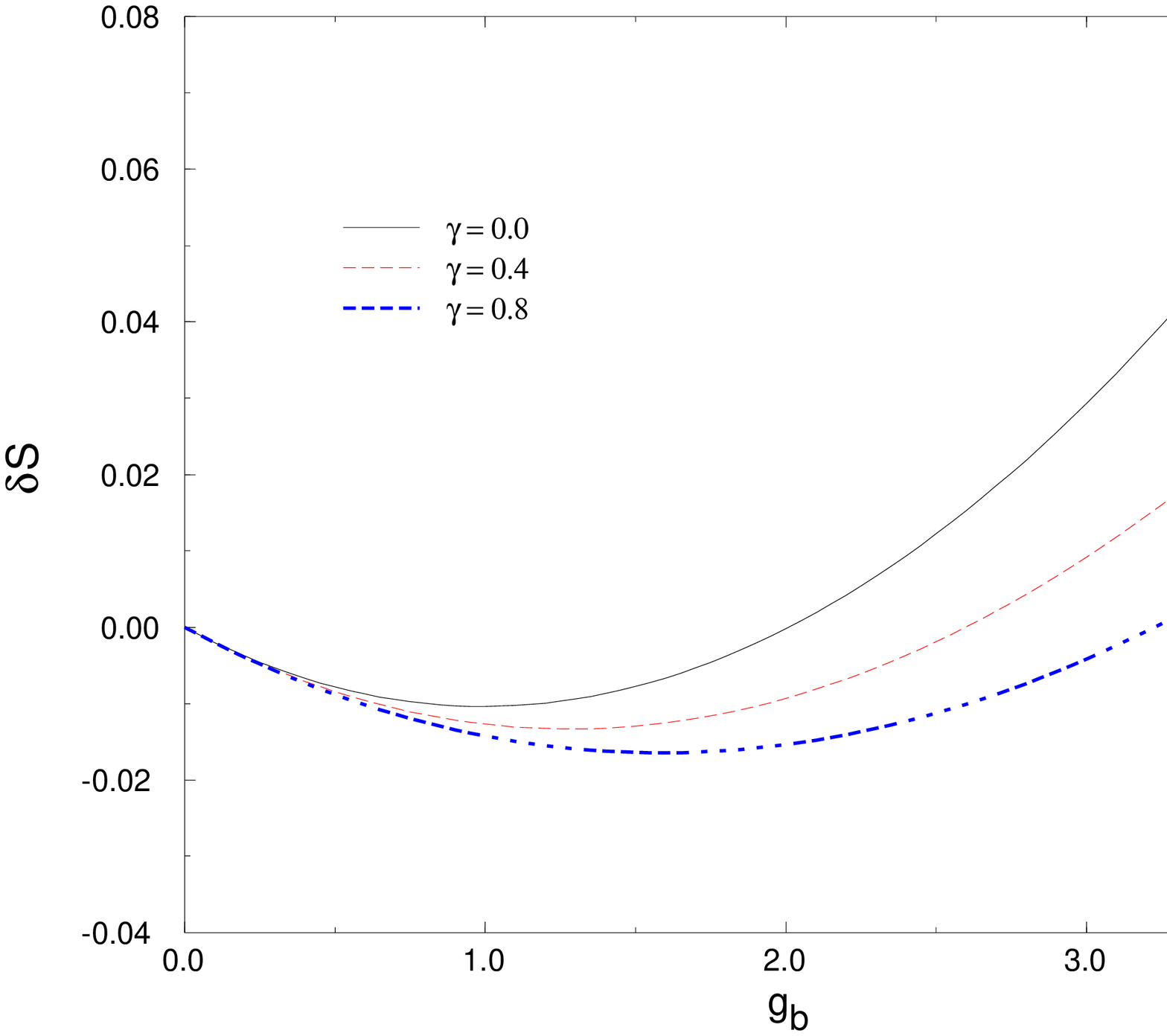,width=8 cm,height=6cm}}
\caption{Results showing dependence of $\delta S$ on $g_{b}$ 
,for $\Gamma = 0.1, l_1 = 0.4, \Delta = 10,\gamma=0,0.4,0.8,$ and $V=0$. }
\label{fig4}
\end{figure}

\begin{figure}
\centerline{\psfig{figure=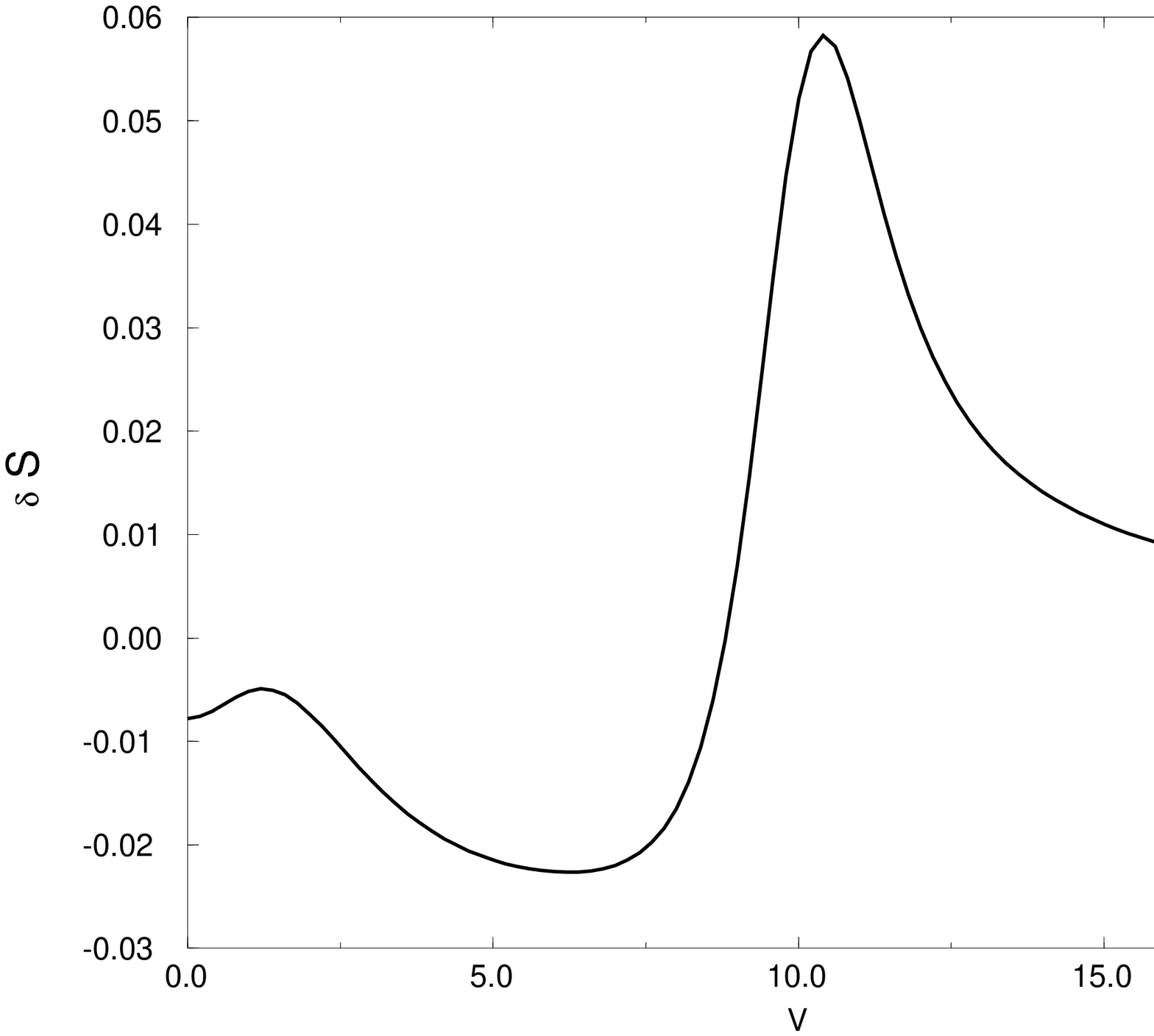,width=8 cm,height=6cm}}
\caption{Results showing dependence of $\delta S$ on $V 
,$for $\Gamma = 0.1, l_1 = 0.4, \Delta = 10,\gamma=0,g_{b}= 1.5,\alpha = 1.0$ . }
\label{fig5}
\end{figure}

\end{document}